\documentclass[fleqn,12pt,twoside]{article}
\usepackage{espcrc1}

\usepackage{graphicx}
\usepackage[figuresright]{rotating}

\newcommand{\AmS}{{\protect\the\textfont2
  A\kern-.1667em\lower.5ex\hbox{M}\kern-.125emS}}

\title{$\pi-\Xi$ correlations in Au-Au collisions at STAR}
\author{Petr Chaloupka  \address[ujf]{Nuclear Physics Institute, Academy of Sciences
        of the Czech Republic, 250 68 Rez near Prague, Czech Republic}
        (for the STAR collaboration) }

\begin{document}
\maketitle

\begin{abstract}
We present $\pi-\Xi$ correlation analysis in Au-Au collisions at \mbox{$\sqrt{s_{{NN}}}=200~GeV$} and \mbox{$\sqrt{s_{NN}}=62.4~GeV$},
performed using the STAR detector at RHIC. A $\Xi^*(1530)$ resonance signal is  
observed for the first time in Au-Au collisions.
Experimental data are compared with theoretical predictions.
The strength of the $\Xi^*$ peak is reproduced in the correlation function 
assuming that $\pi$ and $\Xi$ emerge from a system in collective expansion. 
\end{abstract}

\vspace{-0.1cm}
\section{Introduction}
\vspace{-0.2cm}
Heavy-ion collisions allow the study of strongly interacting matter in
extreme conditions, similar to those of early universe.
At the energy densities achieved in Au-Au 
collisions at RHIC\cite{Gyulassy_QM} the produced matter is expected
to be governed by partonic degrees of freedom.

Current data on spectra and elliptic flow from Au-Au collisions at
\mbox{$\sqrt{s_{NN}}=200~GeV$} \cite{flow_200,v2_200,Fabrice_QM} and
\mbox{$\sqrt{s_{NN}}=130~GeV$} \cite{flow_130,v2_130} suggest that the hot and
dense system created in the collision
builds up substantial collective behavior leading to a rapid transverse 
expansion.
The properties of the induced flow will differ depending on 
whether the collectivity was achieved on a partonic or hadronic
level.
Elliptic flow and spectra data 
of~$\Xi$\cite{chaviere_QM,magali_this} suggest that $\Xi$s pick up 
less flow than other hadrons ($\bar{\Lambda},K,p$).
Because $\Xi$ hadronic cross-section is presumably small, $\Xi$s
are expected to undergo few interactions in the hadronic phase, 
hence picking up less flow and decoupling earlier than $\bar{\Lambda},K,p$.
Therefore $\Xi$s are likely to carry more direct 
information about the partonic stage than other hadrons.

Measurements of final state particle correlations may yield important
insight into the properties of matter created in the earlier
collision times and evolution of the system. Non-identical particle
correlation analyses were suggested\cite{lednicky} to provide
a way of measuring space-time properties
of the particle-emitting source. Compared to standard identical
particle interferometry (sometimes called HBT), which measures one particle species source
size, the non-identical particle correlation function provides a
way to measure relative space-time emission asymmetry among two
particles.

Further insight into the dynamics is provided when studying
the influence of flow on the space-time freeze-out distribution
of the source using a correlation function. Since flow induces
a strong correlation between particle velocities and emission
points, heavier particles are expected to be emitted on average
closer to the surface of the source than the light ones. This leads
to an effective decrease of measured HBT radii and shifts of
average emission points of different particle species relative to
each other \cite{blastvawe}. The study of two particle systems 
with a large mass difference, like $\pi-\Xi$, is therefore of high
interest when studying expansion dynamics. For a detailed discussion
of flow and its effect on particle correlation measurements see
\cite{Fabrice_QM,blastvawe}.

\vspace{-0.1cm}
\section{Data}
\vspace{-0.2cm}
STAR main detector, the Time Projection Chamber, detects charged
particles emerging from primary as well as secondary vertices.
Pions, kaons and protons are identified via their specific energy
loss (dE/dx). We topologically reconstruct
charged $\Xi$ hyperons, which decay via 
$\Xi\rightarrow\Lambda+\pi$, and subsequently
$\Lambda\rightarrow\pi + p$, into proton and two pions.

The $\pi-\Xi$ correlation function $C(k^*)$ was analyzed in the pair
rest frame, where 
\mbox{$\vec{k}^*=\vec{p}^*_{\pi}=-\vec{p}^*_{\Xi}$}.
The distribution of real $\pi-\Xi$ pairs from individual events was
divided by a mixed-event pair distribution, in which each particle
was taken from a different event.
In order to have a well-defined
baseline, events for mixing were divided into individual classes
in multiplicity and primary vertex position. To avoid
fake correlations coming from a mismatched $\Xi$ bachelor pion
(pion from $\Xi\rightarrow\Lambda+\pi$ decay)
we require the momentum difference between the primary and
bachelor pion to be greater than $0.05~GeV/c$ .
For this analysis only primary pions with $|y|<0.5$ were accepted.

Two data sets of Au-Au collisions at energies
\mbox{$\sqrt{s_{{NN}}}=200~GeV$} and \mbox{$\sqrt{s_{NN}}=62.4~GeV$}
were used.  Events recorded at $200~GeV$ consist of two datasets:
the minimum bias sample covers centrality $0\%-80\%$, while the other
set contains $10\%$ most central collisions. The $62.4~GeV$ minimum
bias dataset consists of events with centrality between
$0\%-80\%$.

The correlation function was corrected for pair purity, defined as
the product of the $\pi$ and $\Xi$ purities. The purity of the $\Xi$
sample, was estimated from the signal to combinatoric background ratio
in invariant mass distribution as a function of $p_t$. The pion sample 
purity was estimated from $\lambda$ measured in $\pi$-$\pi$ HBT analysis
\cite{mercedes}.
The influence of the momentum resolution is still under study
and has not been corrected for.

\vspace{-0.1cm}
\section{Results}
\begin{figure}[htb]
\vspace{-0.1cm}
\centering
\includegraphics[width=0.53\textwidth,clip]{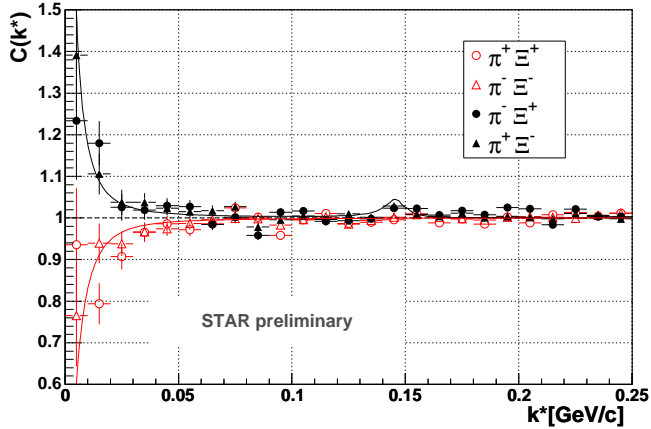}
\vspace{-0.7cm}
\caption{
$\pi-\Xi$ correlation function
 for 10\% most central collisions at 200~GeV.
 Triangles and circles are experimental data, solid
lines are theoretical predictions.}
\label{cf_200}
\vspace{-0.4cm}
\end{figure}

\vspace{-0.2cm}
In Figure~\ref{cf_200}\ we present the preliminary results on
$C(k^*)$ for all four combinations of \mbox{$\pi^{\pm}-\Xi^{\pm}$}
pairs from the 10\% most central $200~GeV$ Au-Au collisions.
In the low $k^*$ region ($k^* < 0.05~GeV/c$) the correlation function is
dominated by the Coulomb interaction. Since $C(k^*)$ is,
within statistical errors, the same for
\mbox{$\pi^{+}-\Xi^{-}$},\mbox{$\pi^{-}-\Xi^{+}$} and
\mbox{$\pi^{+}-\Xi^{+}$},\mbox{$\pi^{-}-\Xi^{-}$} pairs,
correspondingly, the data with like and unlike-sign pairs are
combined, in order to improve the statistics.

\begin{figure}[h]
\begin{minipage}[hl]{0.56\textwidth}
\includegraphics[width=7.73cm, clip]{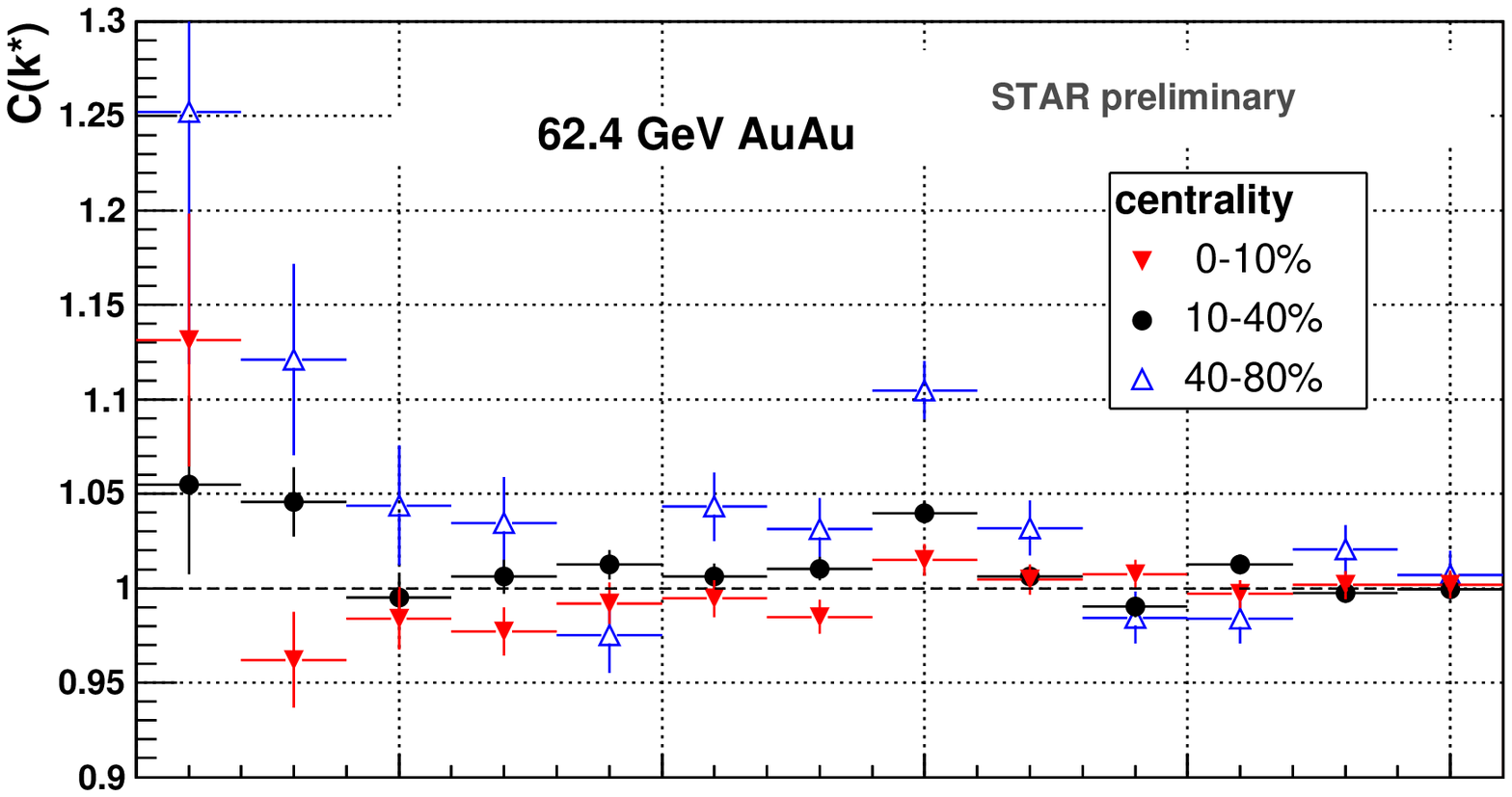}
\includegraphics[width=7.73cm, clip]{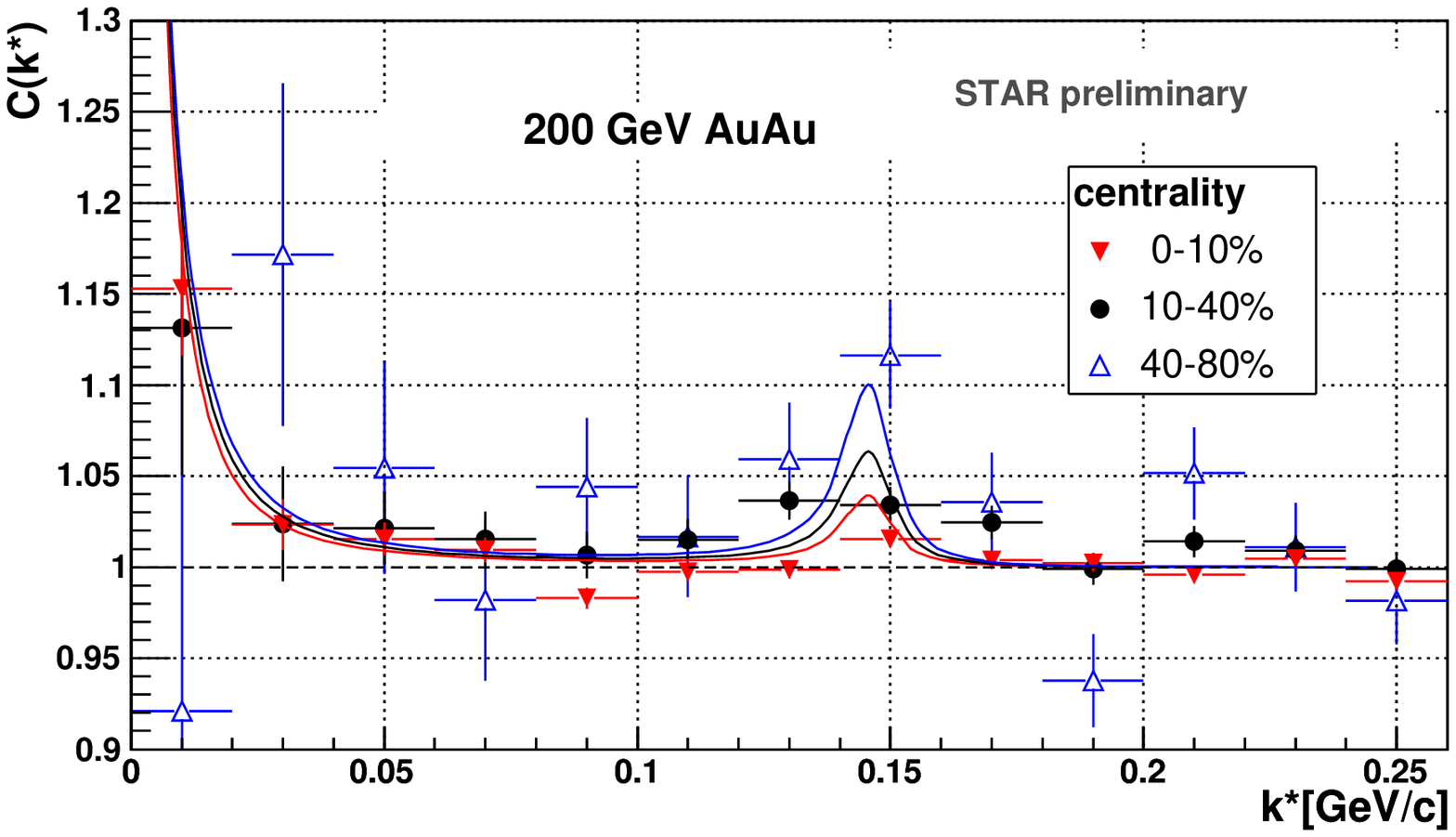}
\vspace{-0.7cm}
\caption{Centrality dependence of $C(k^*)$ for unlike-sign $\pi-\Xi$ pairs;
  upper panel: at 62.4~GeV, lower panel: at 200~GeV.
  Solid lines show theoretical predictions for corresponding centrality bins.}
\label{ccent}
\end{minipage}
\hfill
\begin{minipage}[hr]{0.4\textwidth}
\includegraphics[height=8.6cm]{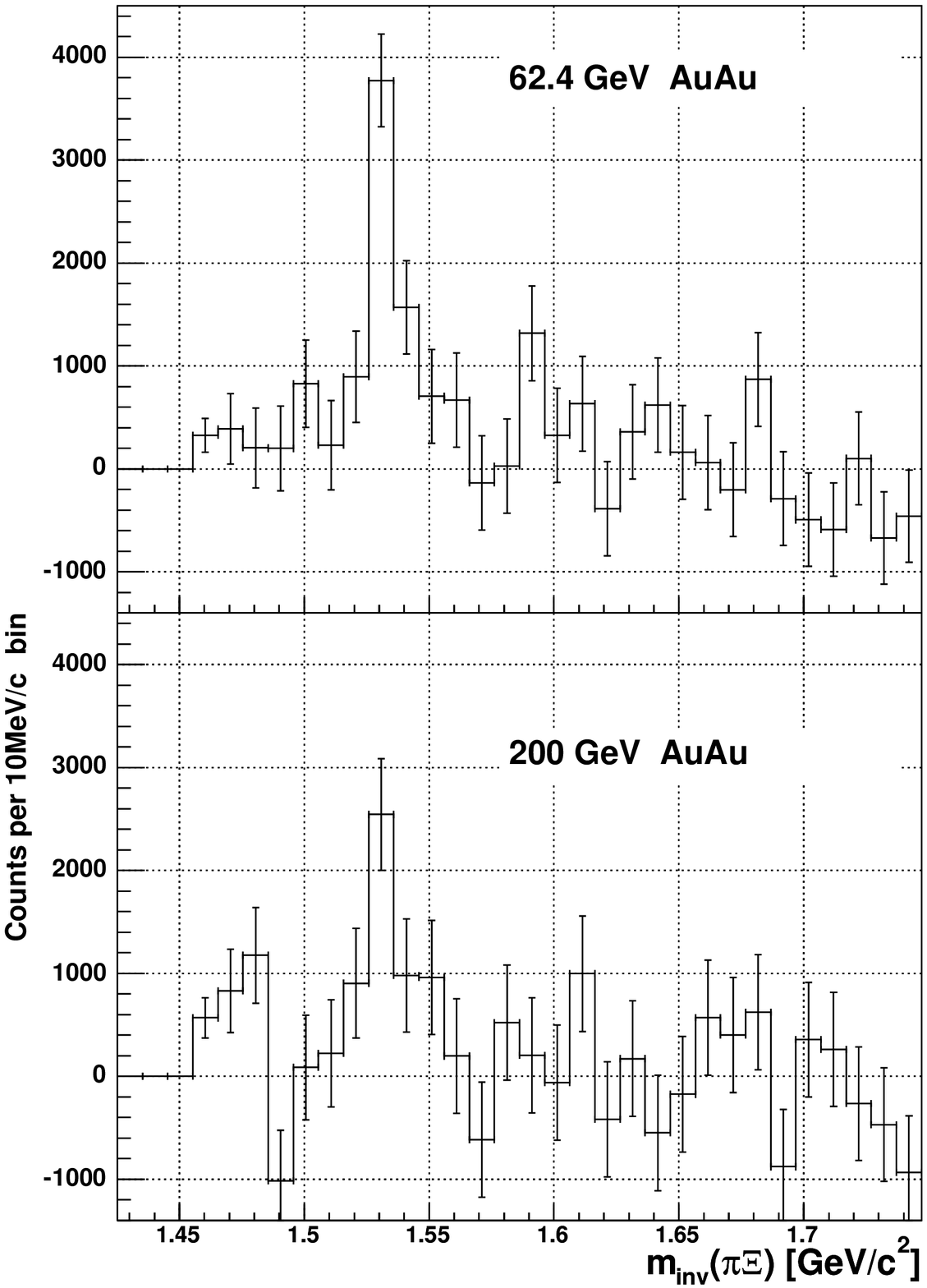}
\vspace{-0.9cm} 
\caption{$\pi - \Xi$ invariant mass distribution of unlike-sign pairs; 
         upper panel: at \mbox{62.4~GeV}, 
         lower panel: at \mbox{200~GeV}.}
\label{minv}
 
\end{minipage}
\vspace{-0.4cm}
\end{figure}

In Figure~\ref{ccent}\ the centrality dependence of $C(k^*)$ for
unlike-sign pairs at $62.4~GeV$ and $200~GeV$ is presented. For
statistics reasons only three centrality bins are used. Even
though the analysis is statistically challenging,
the correlation functions exhibit the same general features
at all centrality bins and energies.
The Coulomb region is mostly affected by the low statistics which
does not make it possible to observe conclusively any centrality
dependence. However a peak at  $k^{*}\approx 0.15~GeV/c$,
corresponding to \mbox{$\Xi^*(1530)\rightarrow \Xi + \pi$} decay is
clearly visible. $k^*$ is directly connected to the
invariant parent particle mass $M$:\mbox{$k^*=\sqrt{[
M^2-(m_1-m_2)^2]~[M^2-(m_1+m_2)^2]}/2M$}. As a crosscheck
we verify that the $\Xi^*$ peak is also visible in the invariant mass
distribution as shown in Figure~\ref{minv}. 
The distributions show a visible $\Xi^*$ peak in both datasets.
No extraction of the yield has been attempted so far.
Most importantly, the $\Xi^*$ peak in
$C(k^*)$  shows significant centrality dependence.
 
\vspace{-0.1cm}
\section{Theory comparison}
\vspace{-0.2cm}
The correlation function $C(\vec{k}^*)$ of two particles with
momenta $\vec{p}_a$ and $\vec{p}_b$ is determined by source properties,
characterized by  $g(\vec{r}^*)$, the normalized probability of
two particles' emission points being separated by a distance
$\vec{r}^*$ in the pair's center-of-mass frame.
 \mbox{$C(\vec{k}^*)=\int{ g(\vec{r}^*) |\psi(\vec{k}^*,\vec{r}^*)|^2  \,d^3{r^*}}$} ,
where $\vec{k}^*=\vec{p}^*_a=-\vec{p}^*_b$ is the particle momentum
in the pair rest frame.

Thus knowledge of the wave function $\psi(\vec{k}^*,\vec{r}^*)$ of
two interacting particles, allows to predict $C(\vec{k}^*)$.
A method to evaluate $\psi(\vec{k}^*,\vec{r}^*)$ for different
systems of non-identical particles (including $\pi-\Xi$ system),
incorporating the Coulomb and strong interactions has been
recently proposed in \cite{pratt_model}
(generalizing the method in \cite{lednicky2}).

Theoretical predictions for $C(|\vec{k}^*|)$ are shown in
Figure~\ref{cf_200}\ and Figure~\ref{ccent}\ together with corresponding
real data.
Standard $R_{out},R_{side},R_{long}$
source parameterization\cite{bertsch_pratt} has been used for
both $\pi$ and $\Xi$ sources.
The values were obtained from
hydro-inspired blast wave calculations \cite{blastvawe}, with the 
assumption that $\Xi$s flow as pions, kaons and protons .
The parameters for blast wave fit were obtained from fitting spectra\cite{flow_200}
and $\pi$-$\pi$ HBT radii\cite{mercedes}
extracted from 200GeV Au-Au STAR data. 
The parameters used in he blast wave fit for different centralities
can be found in \cite{mercedes}.

Similarly to experimental data the  theoretical $C(k^*)$ of unlike-sign pairs
shows peak corresponding to $\Xi^*(1530)$ resonance. Moreover the
peak strength exhibits significant sensitivity to the size of the
source.

\vspace{-0.1cm}
\section{Conclusions}
\vspace{-0.2cm}
We have presented first measurements of the $\pi-\Xi$ correlation function
$C(k^*)$ in heavy ion collisions. The results, though still preliminary,
show clearly the effects of Coulomb and strong interactions
among produced $\pi$ and $\Xi$.  In particular, strong interactions leading
to the $\Xi^*(1530)$ resonance signal is observed for the first time in
heavy-ion collisions. Moreover, in the region of $\Xi^*(1530)$, the
$C(k^*)$ shows much stronger sensitivity to the source size than in the
Coulomb region.

The theoretical prediction, under assumption of both particles 
exhibiting transverse radial flow, reproduces experimental data
within statistical errors.
However, to unambiguously conclude on $\Xi$ flow, 
it is necessary to probe shifts of average space-time emission points
constructing $C_{+}(k^{*})/C_{-}(k^{*})$ ratio\cite{lednicky}.
We expect that sufficient statistics to perform this study 
will be available from RHIC 2004 data run. 

\vspace{-0.3cm}
\section*{Acknowledgements}
\vspace{-0.25cm}
The author wishes to thank S.~Pratt for making 
his code of model calculations  available to us.
This work was supported by the Grant Agency of the Czech Republic grant 202/04/0973.

\end{document}